\documentclass[a4paper,twocolumn,showpacs,showpacs,superscriptaddress,prb,floatfix]{revtex4}
\usepackage{graphicx,placeins}          
\usepackage{dcolumn}
\usepackage{amssymb,amsmath}
\usepackage{color}
\usepackage{epsfig}

\begin{document}

\title{Polarisation of THz synchrotron radiation: from its measurement to control}

\author{Meguya Ryu}
\affiliation{Tokyo Institute of Technology, Meguro-ku, Tokyo
152-8550, Japan}
\author{Denver Linklater}
\affiliation{Swinburne University of Technology, John st.,
Hawthorn, Victoria 3122, Australia}
\author{William Hart}
\affiliation{Swinburne University of Technology, John st.,
Hawthorn, Victoria 3122, Australia}
\author{Armandas Bal\v{c}ytis}
\affiliation{Swinburne University of Technology, John st.,
Hawthorn, Victoria  3122, Australia}
\author{Edvinas Skliutas}
\affiliation{Laser Research Center, Department of Quantum
Electronics, Faculty of Physics, Vilnius University,
Saul\.{e}tekio Ave. 10, Vilnius LT-10223, Lithuania}
\author{\\\protect Mangirdas Malinauskas}
\affiliation{Laser Research Center, Department of Quantum
Electronics, Faculty of Physics, Vilnius University,
Saul\.{e}tekio Ave. 10, Vilnius LT-10223, Lithuania}
\author{Dominique Appadoo}
\affiliation{Infrared Microspectroscopy Beamline, Australian
Synchrotron, Clayton, Victoria 3168, Australia}
\author{Yaw-Ren Eugene Tan}
\affiliation{Australian Synchrotron, Clayton, Victoria 3168,
Australia}
\author{Junko Morikawa}
\affiliation{Tokyo Institute of Technology, Meguro-ku, Tokyo
152-8550, Japan}\email{morikawa.j.aa@m.titech.ac.jp}
\author{Saulius Juodkazis}
\affiliation{Swinburne University of Technology, John st.,
Hawthorn, Victoria  3122, Australia} \affiliation{Melbourne Center
for Nanofabrication, Australian National Fabrication
Facility\\Clayton,Victoria~3168, Australia}

\begin{abstract}
Polarisation analysis of synchrotron THz radiation was carried out
with a standard stretched polyethylene polariser and revealed that
the linearly polarised (horizontal) component  contributes up to
$22\pm 5\%$ to the circular polarised synchrotron emission
extracted by a gold-coated mirror with a horizontal slit inserted
near a bending magnet edge. Comparison with theoretical
predictions shows a qualitative match with dominance of the edge
radiation. Grid polarisers 3D-printed out of commercial acrilic
resin were tested for the polariser function and showed spectral
regions where the dichroic ratio $D_R
> 1$ and $<1$ implying importance of molecular and/or stress
induced anisotropy. Metal-coated 3D-printed THz optical elements
can find a range of applications in intensity and polarisation
control of THz beams.
\end{abstract}

\date{\today}

\pacs{polarisation,  FT-IR, synchrotron, anisotropy of absorbance,
3D printing}
\maketitle
\section{Introduction}

Methods of terahertz generation are evolving fast using
ultra-short laser pulses, electrically driven 2D electron gas in
semiconductor junctions, and photo-mixing, laser-driven electron
plasmas to facilitate number of applications which require smaller
and portable
devices~\cite{HuangO,Carr,Hafez,Tian,Shur,Nanni,17apl202101}. The
highest intensity THz sources are available at  free electron
laser and synchrotron facilities~\cite{Tan,Knyazev}. With a
high-brilliance synchrotron THz radiation it is possible to use
imaging arrays and monitor \emph{in situ} temporal evolution,
e.g., of a phase transition in real time. Polarisation of
synchrotron THz radiation becomes important in absorbance
spectroscopy and for investigation of orientational
anisotropy~\cite{17m356,17sr-silk}. Even for non-absorbing
transparent materials at THz spectral range, a birefringence would
cause polarisation changes due to a purely refractive phase delay
which is important for interpretation of spectroscopical data.
Synchrotron THz wavelengths are spanning wavenumber range
7-700~cm$^{-1}$  (or 0.2 - 21~THz), which in terms of wavelengths
are 1.43~mm - 14.3~$\mu$m. With a widely accessible 3D printing
technology reaching tens-of-micrometers resolution of plastic
components, there is a potential to make optical elements for THz
applications. Rough surfaces on $\sim 5 - 10~\mu$m scale on 3D
printed surfaces can be smoothed by controlled reflow using
strongly absorbed UV post-illumination of the plastic
workpieces~\cite{Chidambaram} and are reaching required
sub-wavelength $\lambda/10$ min-max roughness even for the
visible-IR spectral range applications. Plastics have refractive
index of $\sim 1.5$ at far-IR - to - THz spectral range with a
variable absorbance, which has to be small for highly efficient
optical elements~\cite{Cunningham}. Molecular ordering in polymers
cause birefringence and an anisotropy of absorbance, which can be
determined by the four polarisation method~\cite{Hikima} used in
this study.

\begin{figure}[tb]
\begin{center}
\includegraphics[width=8.50cm]{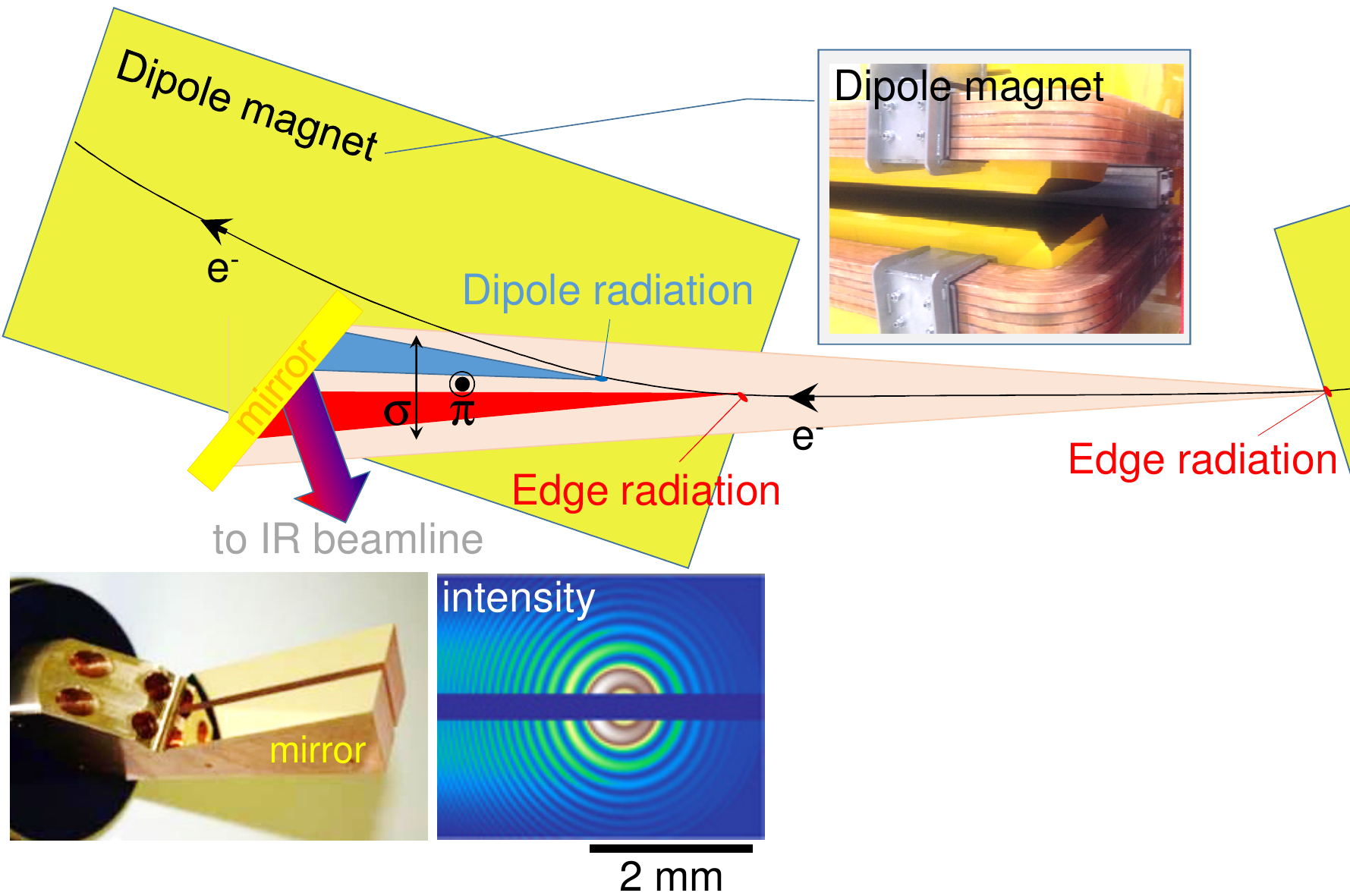}
\caption{Source of dipole and edge radiation which both have
$\sigma$ (horizontal along the slit) and $\pi$ (vertical)
components collected for the use at the IR beamline. Further along
the beamline there is a beam splitter box that then directs the
dipole radiation to the mid-IR branch of the beamline and the edge
radiation to the far-IR branch. Pickup of THz radiation is made
with the first mirror (photo). Intensity shows distribution of
IR-THz radiation at the first mirror reflection. }
\label{f-dipole_edge_radiation}
\end{center}
\end{figure}

Polarisation of synchrotron THz radiation has a complex structure
which has to be well understood when beam is focused (or imaged)
onto the sample with spot sizes down to sub-1~mm from the first
mirror where the beam has cross sections of $\sim 2$~cm. The
synchrotron THz radiation is created by relativistic electrons
traveling through a dipole magnet.

The observed radiation from a dipole magnet in the orbital plane
of the electron is referred to as $\sigma$-polarisation mode
radiation and is typically horizontally polarised. In the
perpendicular plane with a non-zero vertical observation angle,
$\psi$, we have the $\pi$-polarisation mode radiation and is
typically vertically polarised. The combination of the $\sigma$
and $\pi$ modes results in elliptically polarised radiation above
and below the deflection plane of the electron beam
(Fig.~\ref{f-dipole_edge_radiation}). The proportion of the
integrated power of the $\sigma$ mode to $\pi$ mode radiation is,
$P_{\sigma} = \frac{7}{8}P_{tot}$ and $P_{\pi} =
\frac{1}{8}P_{tot}$ for typical dipole radiation~\cite{Balerna}.

The THz spectrum of the synchrotron radiation used by the far-IR
branch of the IR beamline at the Australian Synchrotron (AS) has a
horizontal acceptance of -14~mrad to +14~mrad and focuses on using
the radiation created by the electron beam just as it exits and
enters the magnetic field of the dipole
magnet~\cite{chubar1993generation,chubar1993vuv}. This is referred
to as edge radiation. The spatial and wavelength dependent
distribution of the linearly and elliptically polarised radiation
is more complex at the edge than in the body of the dipole
magnet~\cite{Balerna,chubar1995precise} and is discussed in
Sec.~\ref{append_edge}.

In THz spectral range, it is possible to use linear polarisers to
set polarisation at the sample with metallic grid or metallic
gratings with high extinction ratio defined by transmission for
the transverse magnetic and electric modes $E_r = T_{TM}/T_{TE}
\simeq 45$~dB at 1~THz~\cite{Yamada}. This is achieved at the cost
of a reduced intensity. By phase and polarisation control with
3D-printed optical elements with small absorption losses, it would
be possible to create more efficient optical elements for
polarisation control.

Here, we report results of a polarisation study of THz synchrotron
radiation and characterisation of 3D-printed acrilic grid
waveplate polarisers. Simulation of polarisation at the bending
magnet edge is compared with experimental measurements at the
sample location.
\begin{figure}[tb]
\begin{center}
\includegraphics[width=6.50cm]{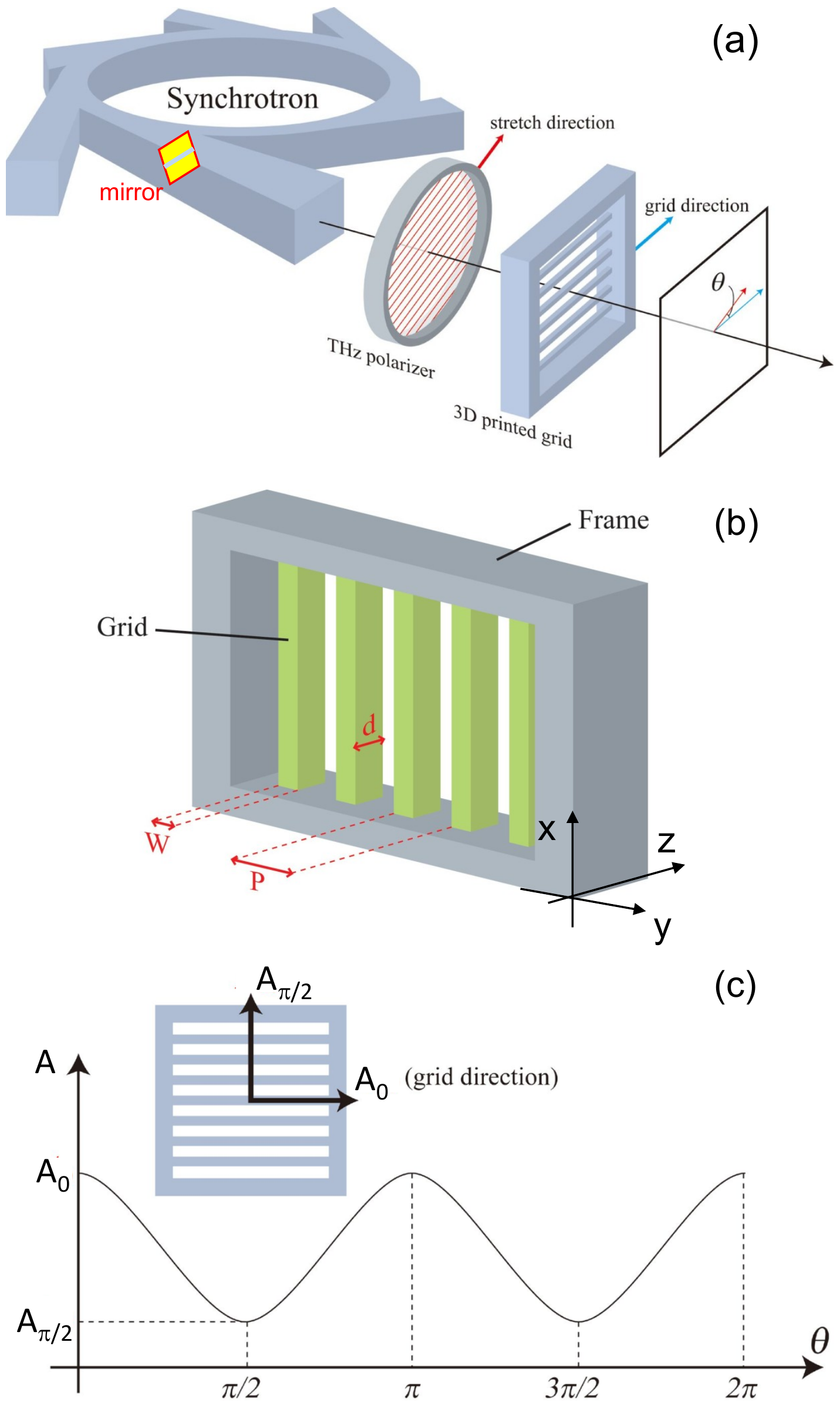} \caption{(a) Geometry of experiment designed to characterise polarisation of the THz radiation by mirror with 3-mm-wide mid-gap.
(b) 3D printed grating used as polarisers. (c) Expected angular
dependence of the absorbance, $A = -\lg T$, from sample with
orientational anisotropy.} \label{f-exper}
\end{center}
\end{figure}
\begin{figure}[tb]
\begin{center}
\includegraphics[width=8.50cm]{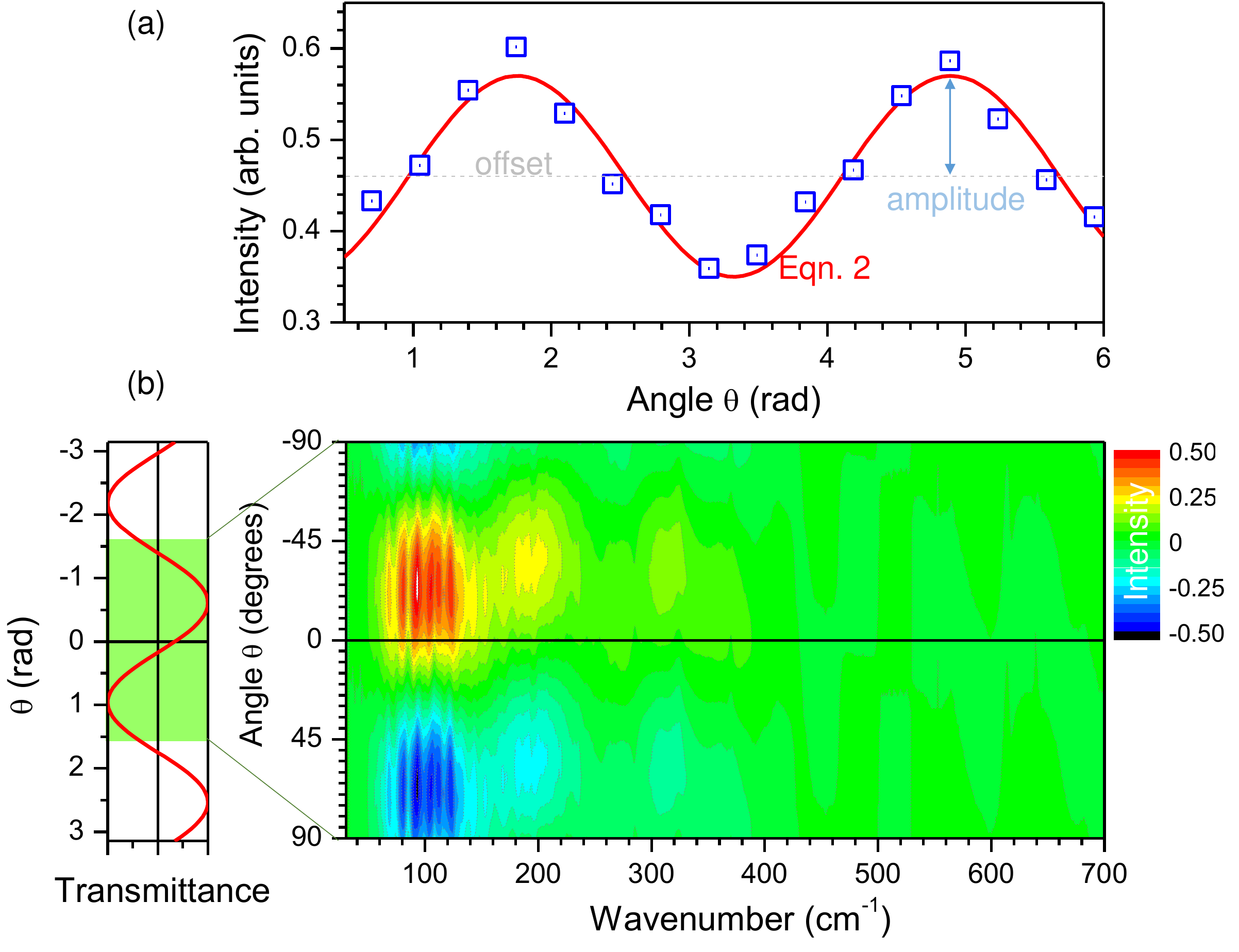} \caption{(a) Experimental orientational dependence of
synchrotron radiation transmission (squares) through the
polyethylene polariser and the best fit by Eqn.~\ref{e1}; the
phase was chosen for the best fit. (b) Spectrum-polarisation map
of synchrotron radiation (the offset subtracted).  } \label{f-pol}
\end{center}
\end{figure}

\begin{figure}[tb]
\begin{center}
\includegraphics[width=8.5cm]{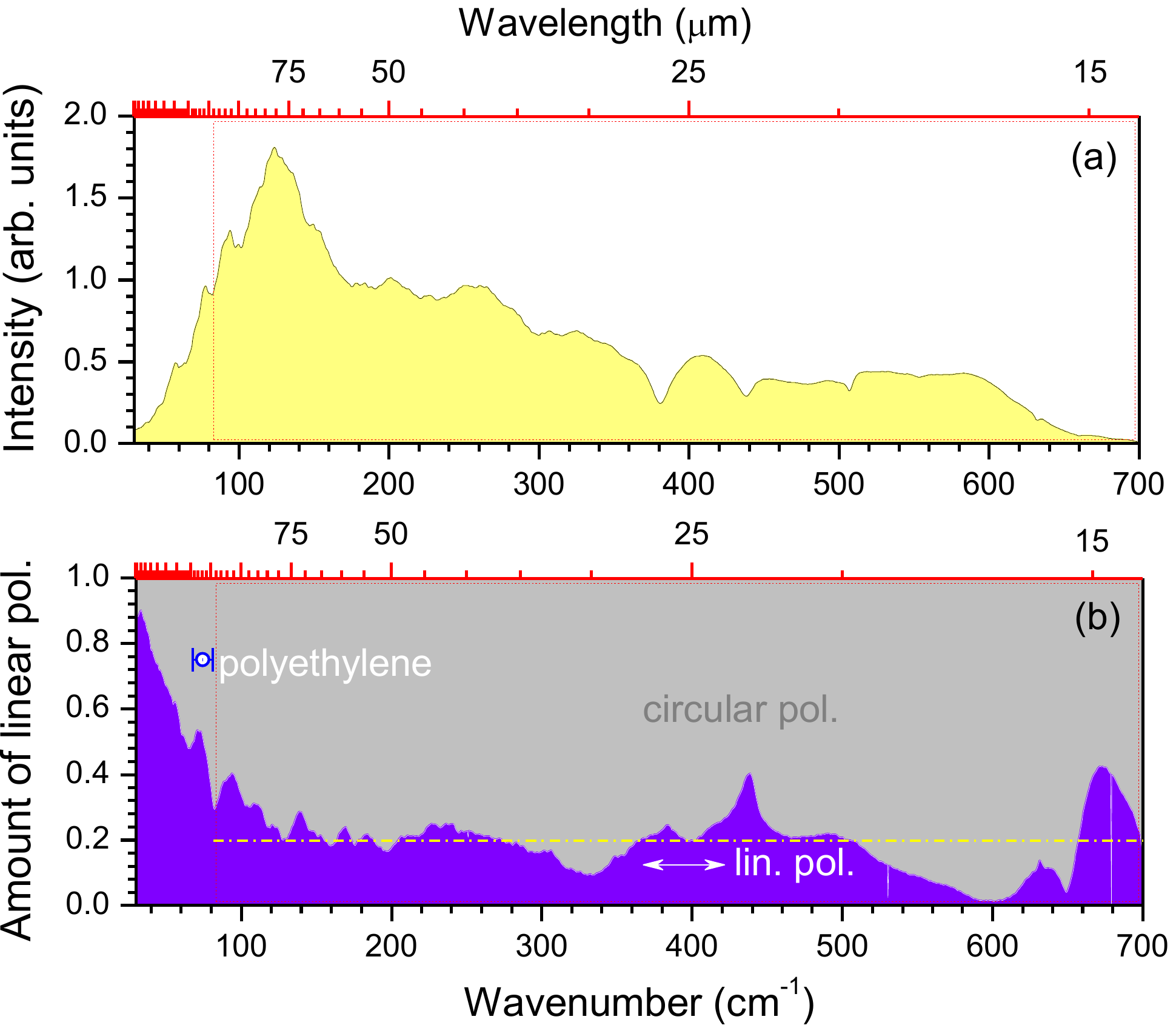} \caption{(a) THz radiation spectrum (unpolarised). (b) Portion of the linearly polarised component $Lin$ (Eqn.~\ref{e2}.) The spectral position of the
absorption band of polyethylene at $2.23\pm 0.2$~THz~\cite{Koch}
is shown by the circular marker. Horizontal line at $Lin \simeq
0.22$ defines the average contribution of the linearly polarised
(horizontal) component of E-field. The region enclosed in the box
is where transmittance of the polyethylene polariser is constant.}
\label{f-lin}
\end{center}
\end{figure}

\section{Methods}

Polarisation analysis  of synchrotron radiation was carried out at
the IR Beamline on Australian Synchrotron,
Melbourne~\cite{Tan1,Einfeld} with a standard linear stretched
polyethylene (PE)  polariser (Bruker No. F251; 45$-\mu$m-thick
PE). Figure~\ref{f-exper} shows schematics of experiments for
polarisation analysis and absorbance anisotropy measurements of
the 3D-printed micro-gratings.

\subsection{3D-printed THz optical elements}

Plastic gratings of varying aspect ratio (the depth-to-width) $AR
= d/w$, duty cycle $DR = w/P$ and period $P$ (see geometry in
Fig.~\ref{f-exper}(b)) were fabricated using the Ember 3D printer
(AutoDesk). The Ember printer possesses a 405~nm wavelength light
emitting diode (LED) source and a Texas Instruments digital
micro-mirror (DMD) projection system which facilitates the
UV-curing of an entire $s = (10-100)~\mu$m layer over a single
exposure. The Ember 3D printer is capable of an xy-resolution of
50~$\mu$m, and a z-resolution of 10-100~$\mu$m. Gratings were set
in a frame to allow them to be fixed to a polarisation filter
mount. Grating width $w$ ranged between 100-200~$\mu$m, period $P$
varied between 100-400~$\mu$m, aspect ratio ranged from 1 to 4 and
duty cycle was adjusted between 25\%, 33\% and 50\%.

Commercially available MakerJuice G+ (red) and SF (green) resins
were obtained from MakerJuice Labs, US. They are UV-curable
acrylate-based resins with low viscosity and fast-curing
capabilities. Also, a proprietary Ember resin (black) was used as
supplied or was mixed from constituent ingredients, however, grids
were less well developed due to a non-resolved parts of the
unexposed segments.

Design files and stereo-lithography STL model files were generated
using OpenSCAD (www.openscard.org), a parametric CAD program. A
custom C\# application was used to automatically generate a range
of design files and STL model files within the design space. To
print gratings with a large phase retardance or absorbance along
propagation of the THz beam (z-axis; Fig.~\ref{f-exper}(b)), the
printing sequence followed stacked exposures as samples are moved
along the x-axis. Using Autodesk software, Print Studio, sample
files were oriented perpendicular to the build-head to allow
sequential stacking of model layers, ``growing'' the grating in a
layer by layer mode as the build-head is rotated across the resin
tray, and lifted step-wise after the exposure of each model layer.
Gratings were printed in 25~$\mu$m layers, with a 5~s exposure
time for 4 burn-in layers, 8~s exposure for the first layer and
1.4~s exposure per each subsequent 25~$\mu$m layer. The print
speed of 25~$\mu$m layers was 18~mm/h. The build head was
optionally covered in Kapton tape to assist removal of the
samples. 3D printed gratings were detached from the build head,
then rinsed in a sequence of acetone, isopropanol and water to
remove uncured resin.

For comparison of polariser performance, 3D-printed grids were
coated with a 100~nm sputtered gold film. Such gratings can be
considered having no transmission in the beam region for the light
polarised along $\theta = 0$ azimuth (a strong extinction due to
reflection and absorption) and simulates performance of the
gratings with $AR = \infty$.

\subsection{Optical characterisation}

The dichroic ratio is defined by the maximum-to-minimum absorbance
ratio $D_r \equiv A_{0}/A_{\pi/2}$, where $A_{0,\pi/2}$ are
absorbance values at two perpendicular orientation angles
$\theta$. The $D_r$ was determined for differently prepared
3D-printed grid polarisers; see, the plot is for the absorbance,
$A$, in Fig.~\ref{f-exper}(c). When $D_r = 1$ material (pattern)
has isotropic absorbance while the metallic grid polariser has
$D_r
> 1$ as the E-field of light polarised along the metallic lines of
the grid is absorbed while the absorbance $A_{\pi/2}$ is smaller.
The case of $D_r < 1$ would correspond to an anisotropy induced by
a particular molecular alignment or internal stress in a
grating-like pattern. The anisotropy of absorbance can be
determined from the angular dependence of $A_\theta$ and only four
angles with angular separation of $\pi/2$ are required to retrieve
the fit as shown in Fig.~\ref{f-exper}(c)~\cite{Hikima}:
\begin{equation}\label{e-4a}
A_\theta = A_0\cos^2(\theta) + A_{\pi/2}\sin^2(\theta).
\end{equation}

\begin{figure}[tb]
\begin{center}
\includegraphics[width=8.50cm]{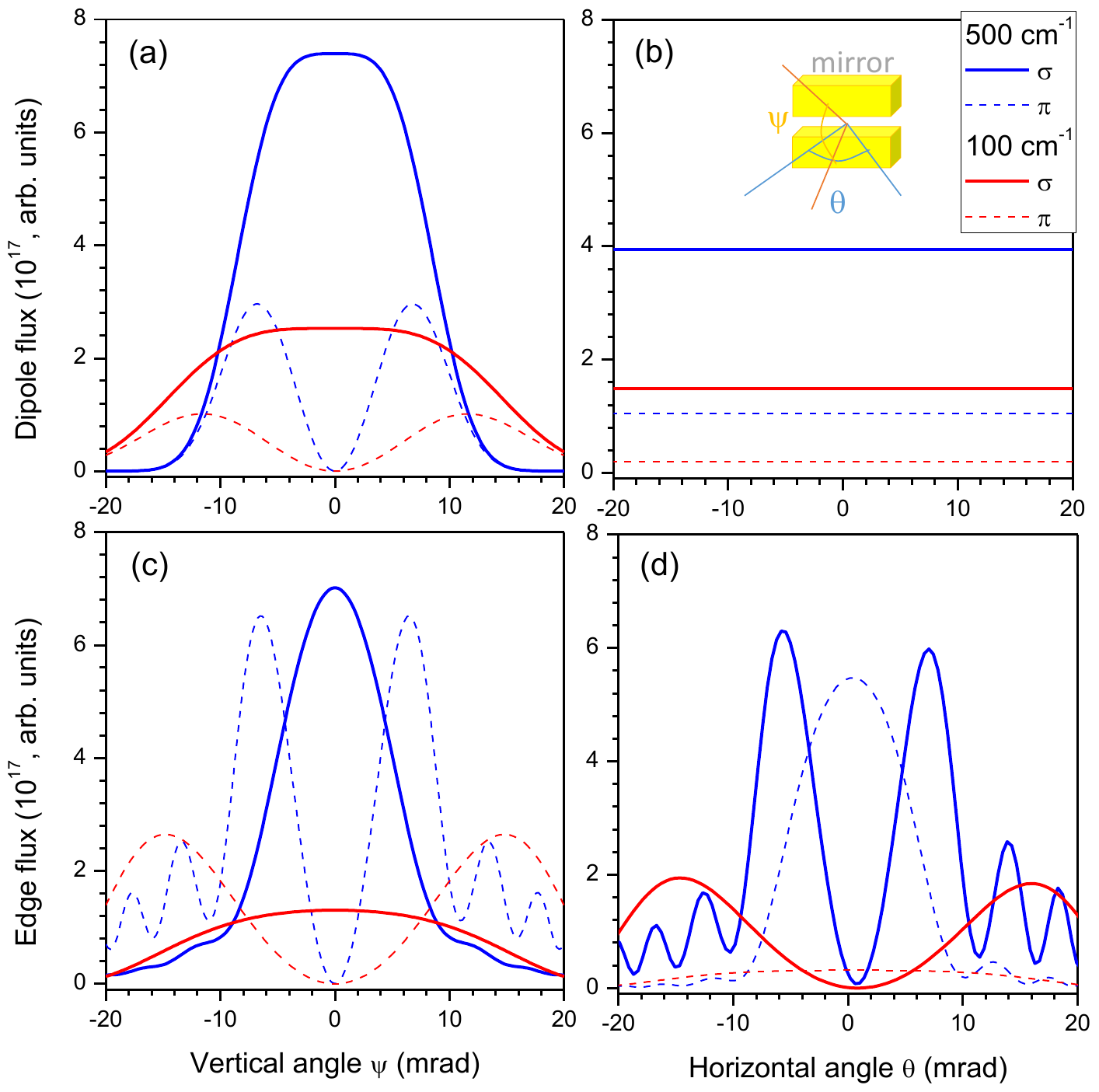}
\caption{(a) and (c) Compares the $\sigma$ and $\pi$ modes for
100~cm$^{-1}$ and 500 cm$^{-1}$ as a function of the vertical
observation angle, $\psi$ while (b) and (d) shows the change a
function of the horizontal observation angle (in the plane of the
electron beam), $\theta$. The comparison highlights the complex
nature of the edge radiation. Inset in (b) schematically shows the
first mirror and angular spread of the IR-THz beam extracted into
the beamline. } \label{f-spatial_distribution}
\end{center}
\end{figure}

This method was recently used to determine anisotrophy of
absorbance of silk at mid-IR spectral range~\cite{17sr-silk}.

\section{Results and Discussion}

\subsection{Polarisation of synchrotron THz radiation}

Synchrotron THz radiation is extracted from the edge of a dipole
magnet using a mirror with a 3-mm-wide slit
(Fig.~\ref{f-dipole_edge_radiation}(b)) to allow higher energy
photons to pass through. The resulting elliptically polarised
radiation can be separated into a circularly polarised (isotropic)
component of the light field $E_I$ and a linear component $E_L$
aligned with the horizontal slit ($\theta = 0^\circ$ azimuth).
Figure~\ref{f-dipole_edge_radiation}(b) shows intensity
distribution of the synchrotron radiation beam taken to the IR
beamline. A standard linear stretched polyethylene polariser was
set into the beam at normal incidence and its orientation,
$\theta$, was changed by a motorized stage while measuring
transmission (Fig.~\ref{f-dipole_edge_radiation}(b)).

\begin{figure}[tb]
\begin{center}
\includegraphics[width=6.8cm]{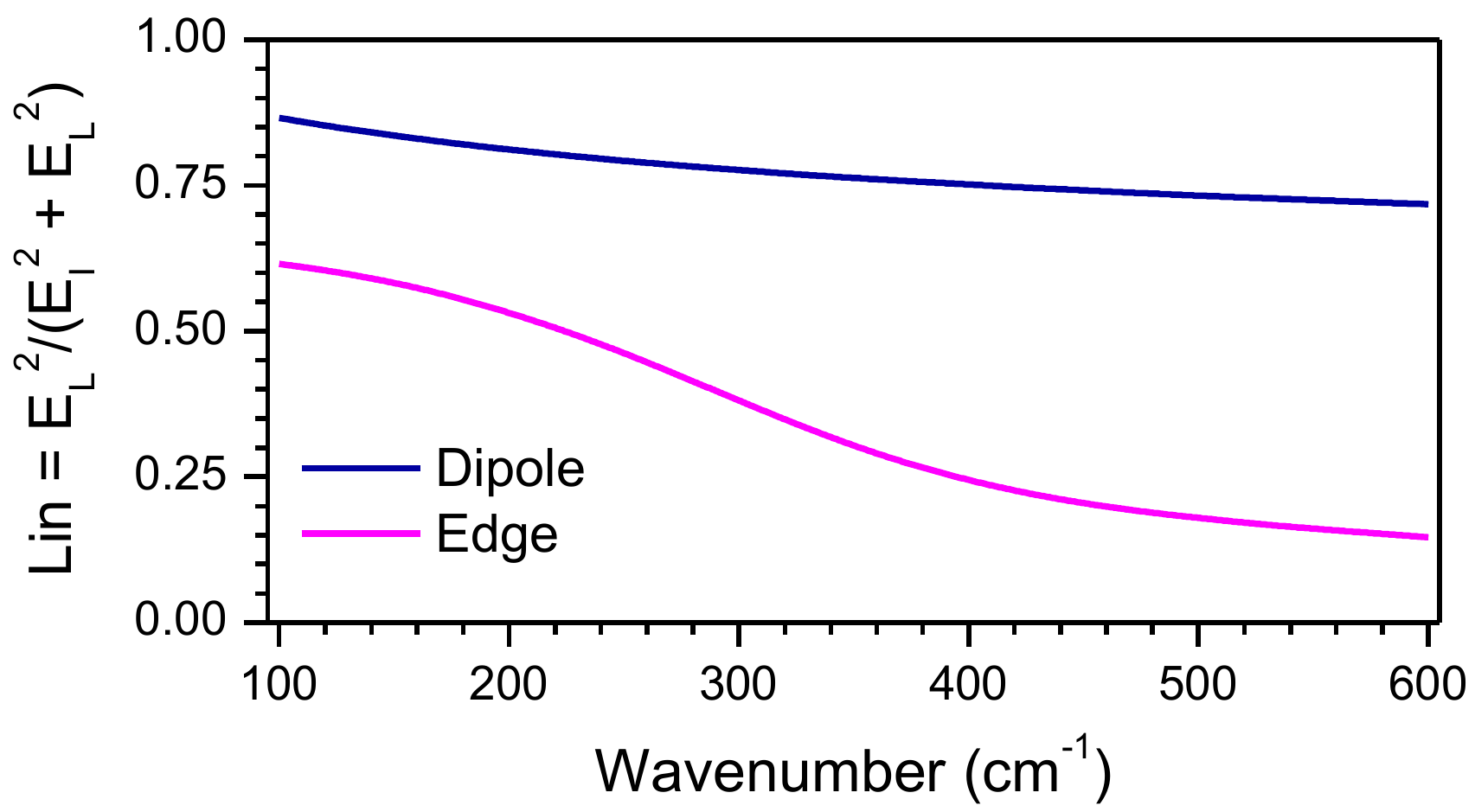}
\caption{Simulation data showing contribution of the linear
polarisation (horizontal; along the first mirror slit,
Eqn.~\ref{e2}); modeling is shown in
Fig.~\ref{f-spatial_distribution} at selected wavenumbers.
Difference in spectral dependencies of $Lin$ ratio shown for the
dipole and edge radiations. Radiation is integrated over the
vertical observation angle, $\psi$ from -8.5 to +8.5~mrad and
horizontally over $\theta$ from -14 to 14~mrad.}\label{f-ratio}
\end{center}
\end{figure}

Figure~\ref{f-pol}(a) shows the angular dependence of
transmission, $T(\theta)$ measured with $20^\circ$ orientation
steps and the best fit to the Malus law $\propto\cos^2(\theta)$
for the analysed case as shown below. For this case of
horizontally polarised linear component $E_L$ ($\theta = 0^\circ$)
mixture with the isotropically (circularly) polarised field,
$E_I$, lets establish an orthogonal base with E-fields at two
azimuths $\theta=0$ and $\theta=\pi/2$, $E_{0,\pi/2}$,
respectively, for the output (transmitted) intensity $E_\theta^2$:
\begin{equation}\label{e1}
\begin{split}
E_\theta^2 & = E_0^2\cos^2(\theta) + E_{\pi/2}^2\sin^2(\theta) \\
& = \left(\frac{1}{2}E_I^2 + E_L^2\right)\cos^2(\theta) +
\frac{1}{2}E_I^2\sin^2(\theta) \\
& = \frac{1}{2}(E_I^2 + E_L^2) + \frac{1}{2}E_L^2\cos(2\theta),
\end{split}
\end{equation}
\noindent where $I_{off} = \frac{1}{2}(E_I^2 + E_L^2)$ is the
offset intensity and $I_{amp} = \frac{1}{2}E_L^2$ is the amplitude
of the transmitted light (through the analyser). The best fit to
$\cos(2\theta)$ dependence according to Eqn.~\ref{e1} provides
$I_{off}$ and $I_{amp}$ values. By measuring angular dependence of
the transmission spectrum, it is possible to present a map which
shows spectrum in abscise and and orientation in the ordinate
directions (Fig.~\ref{f-pol}(b)). It is revealed that the used
polariser has a spectrally broadband action over the entire THz
window (Fig.~\ref{f-pol}(b)).

Exact portion of the linearly polarised light in the entire
spectrum (Fig.~\ref{f-lin}(a)) is estimated by the factor:
\begin{equation}\label{e2}
Lin \equiv \frac{E_L^2}{E_L^2 + E_I^2} = \frac{I_{amp}}{I_{off}},
\end{equation}
\noindent which is plotted in Fig.~\ref{f-lin}(b) and obtained as
the best fit to experimental data at fixed wavelength using
$I_{amp}$ and $I_{off}$ parameters. At the most intense THz
spectral range it was close to 22\%. The highest linearly
polarised intensity is obtained at $\theta = 0^\circ$,
horizontally with the first mirror slit.

\subsection{Edge radiation}\label{append_edge}

A diagram of the source of synchrotron radiation for the IR
beamline is shown in Fig.~\ref{f-dipole_edge_radiation}. The
spatial and spectral distribution of the synchrotron radiation
from a dipole magnet is well understood and using the equations
from ref.~\cite{chao2013handbook} the integrated flux of the
$\sigma$ and $\pi$ polarisations of radiation were simulated.

The dipole radiation is extracted with a rectangular mirror (30
$\times$ 54 $\times$ 460~mm;
Fig.~\ref{f-dipole_edge_radiation}(b)) located $\sim$1.3~m
downstream from the edge of the dipole magnet designed with an
acceptance of 58~mrad in the horizontal plane (-14~mrad to
+44~mrad) and 17~mrad
vertically~\cite{creagh2006design,creagh2007infrared}. Matching of
numerical apertures of the radiation gathering optics along the
entire beamline also plays an important role. The
\emph{\'{e}tendue}, $\varepsilon$, is a measure of the flux
gathering capability of the optical system, i.e., the collected
power is the product of $\varepsilon = area\times solid~angle$
[m$^2$sr] and the radiance of the source [W/m$^2$/sr].

In Fig.~\ref{f-spatial_distribution} the comparison clearly
highlights the complex nature of the edge radiation and is
sensitive to the exact profile of the magnetic field at the edge
of the dipole magnet. These simulations only account for the
distribution of radiation from a single electron, while for exact
simulation it would need to factor in the cross section of the
electron beam
($\sigma_x,\sigma_{x'},\sigma_y,\sigma_{y'}$)~\cite{chubar1995precise}.
Using this single electron model and describing the elliptical
radiation as a combination of linear and circular polarisations
(as presented in Eqn.~\ref{e2}) Fig.~\ref{f-ratio} compares how
the ratio of linear to circular polarisation changes as a function
of wavenumbers for the dipole and edge radiation. A larger
contribution of the horizontal linear polarisation at small
wavenumbers is characteristic for both the dipole and edge
radiation. The measured polarisation ratio $Lin$ at the sample
location (Fig.~\ref{f-lin}(b)) shows a smaller portion of the
linear polarisation due to the \emph{\'{e}tendue}, $\varepsilon$,
and potentially reflects the unmatched numerical apertures for
radiation collection. However, the same trend of larger
contribution of linear (horizontal) polarisation at low
wavenumbers ($< 100$~cm$^{-1}$) as theoretically predicted was
experimentally observed.

\begin{figure}[b]
\begin{center}
\includegraphics[width=8.0cm]{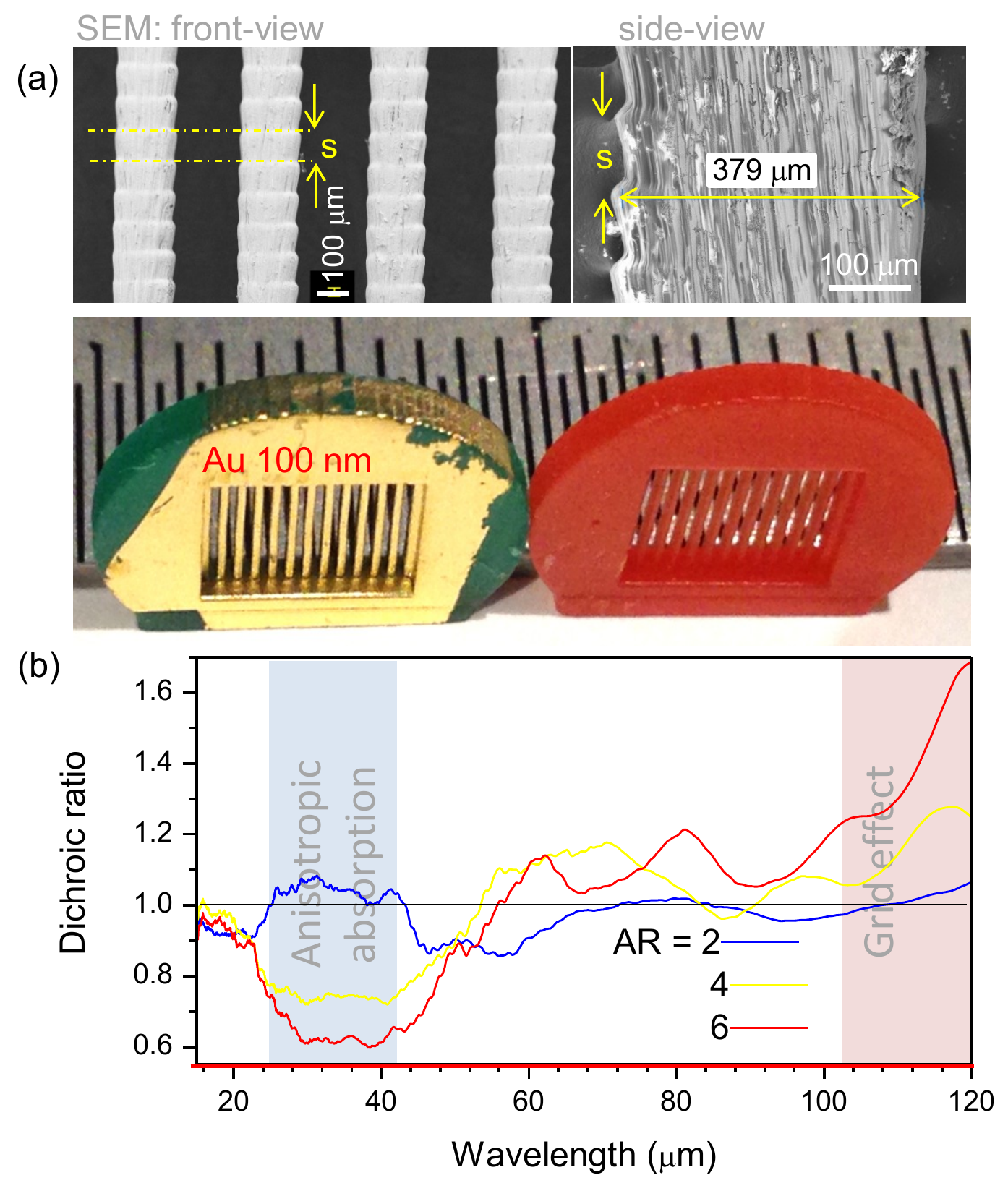}
\caption{(a) SEM front- and side-view images of a 3D-printed
grating and photo images of typical samples made out of MakerJuice
G+ (red) and SF (green) resins; $s$ is the thickness of layer made
in one exposure. Time required to print one sample was
$\sim$5~min; sample is moved upwards with layers added at the
bottom. (b) The dichroic ratio $D_r = A_{max}/A_{min} \propto
T_{min}/T_{max}$ (in other conventions $D_r = A_0/A_{\pi/2} =
A_\parallel/A_\perp$) vs. wavelength for acrylic 3D-printed
gratings of several aspect ratios $AR = \frac{d}{w}$; period $P =
400~\mu$m, duty ratio $DR = \frac{w}{P} = 0.33$. The same spectral
window is enclosed in the box in Fig.~\ref{f-lin}. The color
shaded regions mark spectral ranges where grid effect and
anisotropy of absorption are pronounced.} \label{f-dr}
\end{center}
\end{figure}

\subsection{3D printed polariser grids}

With the fully determined polarisation of the THz radiation,
3D-printed optical elements (Fig.~\ref{f-dr}(a)) were
characterised using the same setup shown schematically in
Fig.~\ref{f-exper} for different duty cycle and aspect ratio
acrylic elements. Figure~\ref{f-dr}(b) shows dependence of the
dichroic ratio $D_r = A_{max}/A_{min} \propto T_{min}/T_{max}$ vs.
wavelength. The $D_r$ was increasing for longer wavelengths as
would be expected for a grid-type polariser where E-field
component along the grid beams is absorbed stronger. This effect
was not very strong since $P
> \lambda$, however, the tendency is clearly recognisable and is more expressed for the higher aspect ratio grids (more
absorbance). As the $AR$ was increasing a tendency of $D_r < 1$
emerged at specific 30-40~$\mu$m wavelength region
(Fig.~\ref{f-dr}(b)). Anisotropy in molecular alignment is an
expected cause (can be linked to the stress in the printed grid).
Similarly,  the speed of silk formation is defining molecular
alignment and mechanical strength of silk
brins~\cite{Shao,Du,Yoshioka}. The optical birefringence and
activity due to an orientational anisotropy is also linked to
anisotropy of absorbance and is one of the consequences of the
molecular alignment~\cite{Chirgadze}.

\begin{figure}[tb]
\begin{center}
\includegraphics[width=7.30cm]{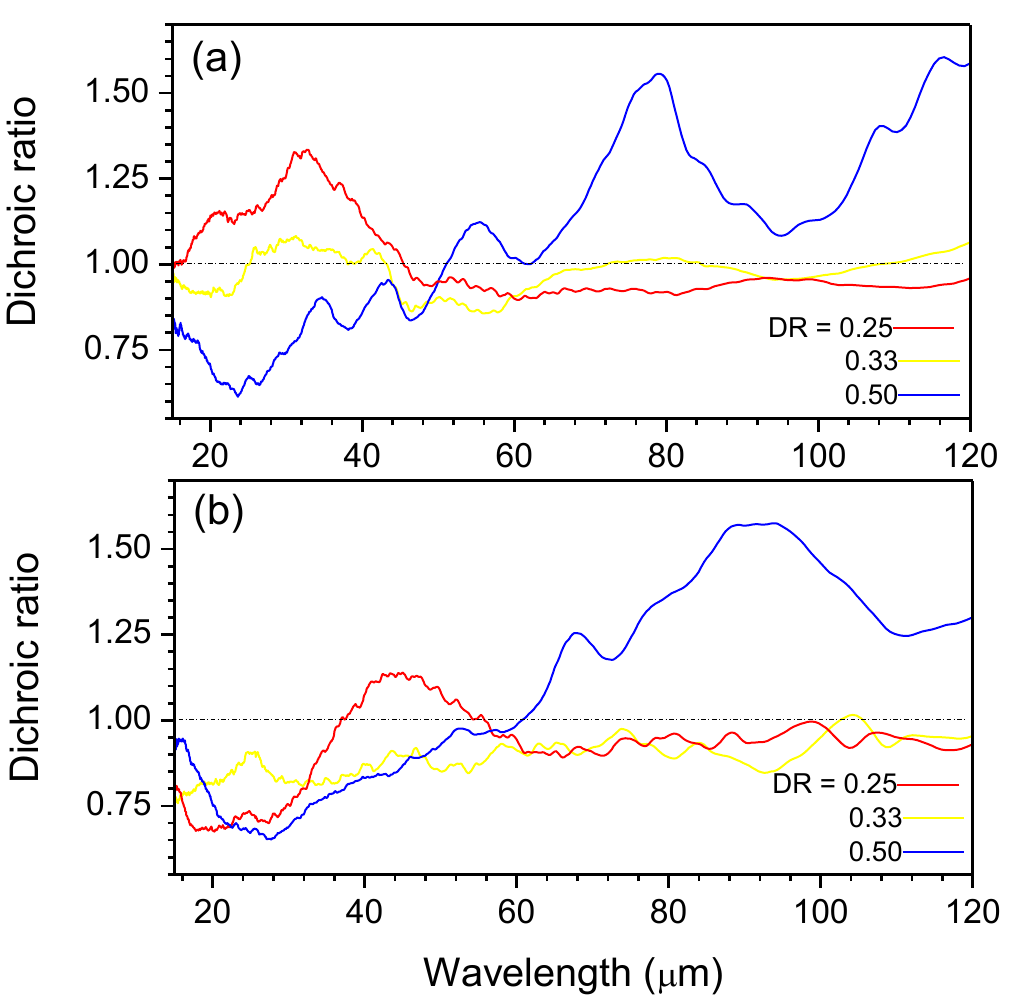} \caption{The dichroic ratio $D_r$ vs. wavelength for acrilic 3D printed grid without (a) and with (b) 100~nm gold coating for several duty ratios
$DR = \frac{w}{P}$; period $P = 400~\mu$m. } \label{f-au}
\end{center}
\end{figure}

Usually, the optimum duty cycle for polarisation optics and phase
retardance $\Delta n\times d$ is at $DR = 0.5$ when sub-wavelength
gratings are fabricated; here $\Delta n$ is birefringence. This
directly follows from the effective medium theory (EMT) which
shows that the smallest height structures required to effectively
phase delay the beam have the minimal height. This usually
corresponds to a desired fabrication condition. Since the EMT can
not be used for the case discussed here with $P > \lambda$, the
efficient phase control cannot be achieved. The dichroic ratio
shown in Fig.~\ref{f-au} is close to $D_r \sim 1$ corresponding to
optically isotropic material for absorption. Only for the largest
duty cycle $DR = 0.5$ an anisotropy begin to  be apparent $D_r
>1$, again as wavelength is increased closer to the EMT range
(Fig.~\ref{f-au}(a)). By sputtering 100~nm of gold, the 3D-printed
grating has to become more anisotropic since THz E-field component
parallel to the grid beams should be absorbed in metal; this can
be considered equivalent to an increase of the aspect ratio
$AR\rightarrow \infty$. However, the effect of $D_r
> 1$ was not strongly pronounced. This can be understood as
being caused by $P > \lambda$.

Dichroic losses measured of a  grating for two linearly polarised
beam orientations $e^{-\Delta^{"}} = \sqrt{P_\parallel/P_\perp}$
provides an indirect measure whether material is suitable for
fabrication of phase elements, optical retarders to control phase,
polarisation, focusing, and orbital momentum. The achievable
efficiency of an optical element, from the material point of view,
is related to the smallest losses defined by the imaginary part of
refractive index $(n^{'}+in^{"})$, which is linked to the
retardance and dichroism by $\Delta = \Delta^{'}+i\Delta^{"}$ with
$\Delta^{(',")}=k[n_{\parallel}^{(',")} - n_{\perp}^{(',")}]d$;
the retardance $\Delta^{'}$ and dichroism $\Delta^{"}$  of the $d$
height optical element (e.g., grating) at wavevector $k =
2\pi/\lambda$ for the wavelength $\lambda$ governs the efficiency
of the optical element. The amplitude of the E-field decreases
exponentially with the hight, i.e., the intensity is given by the
Lambert-Beer's law $I(h) = I_0e^{-2n"\omega h/c} =
I_0e^{-\alpha(\omega)h}$, where $\alpha(\omega) = 2n"k$ is the
absorption coefficient. Then, the amplitude, $Amp$, of the
$\cos$-wave-form measured by the 4-polarisation method
(Eqn.~\ref{e-4a}) is related to the dichroism as:
\begin{equation}\label{e-link}
  Amp \equiv (A_{\parallel} - A_{\perp})/2 = k(n"_\parallel - n"_\perp)h \equiv
\Delta".
\end{equation}
For efficient absorbance and retardance control, 3D printing
technology has to deliver $P < \lambda$ precision of structuring
which begin to be accessible at THz spectral band. Here we showed
that open grid structures with high aspect ratio can be fabricated
at the desired $DR = 0.5$, however, further decrease in the width
of grating beam is required to obtain larger dichroic ratio $D_r$.
The printing method demonstrated here allows 3D printing of
gratings with arbitrary depth, $d$, of the structure.

\section{Conclusions and outlook}

Contributions of the linear and circular polarisation components
in THz beamline spectrum have been determined using polarisation
analysis of transmission.

3D-printed acrylic gratings with rod width of $\sim 100~\mu$m
period, duty cycle 0.5, and aspect ratio up to 8 were made out of
standard acrylic resin. Low absorbance of 3D-printed structures is
promising for fabrication of phase control elements (optical
retarders) which would open new set of polarisation control of THz
beams. The ability to rapidly produce a wide range of complex
optical elements using 3D printing from various materials
(including biocompatible materials) is a key benefit of 3D
printers such as Ember that further enhance this possibility.
Polariser-analyser setup will be required to further investigate
optical retardance due to phase delay in addition to the
absorbance investigated in this study.

Phase control using different approaches tested for visible and
near-IR spectral ranges are transferrable to longer IR and THz
frequencies. Usual phase control relies of defined thickness of
material (the propagation phase)~\cite{10apl211108}, geometrical
phase made by azimuthal patterning of orientation of the optical
axis~\cite{17apl181101}, as well as using metamaterials with phase
control by spectrally overlapping electric and magnetic dipoles in
non-absorbing high-refractive index materials ($n>2$)~\cite{Chong}
which allow to engineer polarisation, intensity, and orbital
angular momentum of the light. The latter two concepts can be
considered as flat optics and can be combined with intensity
control by axicon or Fresnel lens demonstrated recently for THz
wavelengths with performance matching theoretical
efficiency~\cite{Irmantas}. 3D structuring of Si surface by
fs-laser direct oxidation with subsequent plasma etching opens new
possibilities in THz optics where Si is
transparent~\cite{Liu,Sun}.

\subsection*{Acknowledgements}

\small{Experiments were carried out via beamtime project No.~11615
at the Melbourne synchrotron on 8-11 March 2017. This work was
performed in part at the Melbourne Centre for Nanofabrication
(MCN) in the Victorian Node of the Australian National Fabrication
Facility (ANFF). W.H. is supported by an Australian Government
Research Training Program Scholarship. Partial support via NATO
grant SPS-985048 ``Nanostructures for Highly Efficient Infrared
Detection'' is acknowledged. E.S. and M.M. acknowledge financial
support by the OPTIBIOFORM (S-LAT-17-2) project from the Research
Council of Lithuania. J.M. acknowledges the support of JSPS
KAKENHI Grant Number 16K06768. This work was supported in part by
the Global University Project at Tokyo Institute of Technology.}


\bibliographystyle{spiebib}
\small

\end{document}